\documentclass{article}

\usepackage{PRIMEarxiv}

\usepackage[utf8]{inputenc} 
\usepackage[T1]{fontenc}    
\usepackage{hyperref}       
\usepackage{url}            
\usepackage{booktabs}       
\usepackage{amsfonts}       
\usepackage{nicefrac}       
\usepackage{microtype}      
\usepackage{lipsum}
\usepackage{fancyhdr}       
\usepackage{graphicx}       
\usepackage{enumitem}
\usepackage{setspace}
\usepackage{xcolor}

\graphicspath{{Figures/}}     

\makeatletter
\renewcommand\@makefntext[1]{\leftskip=2em\hskip-0.5em\@makefnmark#1}
\makeatother

\title{Use of alternative data: 
High frequency readout of the situation - COVID policies, mobility and R-number
}

\author{
  Ashutosh Mani Dixit \\
  Economist\\
  \texttt{ashutosh.m.dixit@gmail.com} \\
   \And
  Suraj Regmi \\
  Data Scientist \\
  \texttt{regmi125@gmail.com} \\
}

\begin{document}
\maketitle

\begin{abstract}

The role of alternative data in the crisis was recognized even before the COVID-19 pandemic\cite{Hammer_etal_2017}. Now, the months of stalemate made it more urgent to understand the importance of high-frequency data to inform the policy responses \cite{ducharme_james_zhan_2020}. In Nepal, the Government has exerted stay put measures, and physical data collection activities are suspended. The confirmed cases of COVID-19 has reached more than 560,000\cite{mohp} and the country is on high alert .

In this impasse,  the number of secondary cases one would produce over the course of outbreak - the reproduction number ($R_0$) is useful to monitor the transmissibility of COVID-19 \cite{Wallinga_Teunis_2004}. As the R-value is rapidly changing, it can be affected by a range of factors, including not just how infectious a disease is but how Government responds to it, and how the population behaves\footnote{https://www.weforum.org/agenda/2020/05/covid-19-what-is-the-r-number/}. The World Health Organization (WHO) has suggested to the Government of Nepal several recommendations to contain the further spread of COVID-19. To get a sense of how Nepal is coping with the coronavirus pandemic we look at the alternative data sets to get a better understanding of the pandemic policies, mobility, and R-value during COVID.

\end{abstract}


\section{Objective}
\begin{enumerate}[label=(\alph*)]
  \item \textbf{To get the high frequency read out of the COVID situation in Nepal}\\
    We calculate effective reproduction number (R-value) from OWID data (\textit{smoothed})\cite{owidcoronavirus}, and gain additional insights from COVID-19 community mobility reports\footnote{Google community mobility report was launched in April to showcase change in mobility trends in COVID situations.}, the Oxford Coronavirus Government response tracker - Oxford stringency index\footnote{ Policy responses come from the Oxford Coronavirus Government Response Tracker (OxCGRT). The tracker is published by researchers at the Blavatnik School of Government at the University of Oxford\cite{hale2020variation}} and Google search trends.\cite{Google_trends}
  \item \textbf{Make available the source code for extracting alternative data}\\
  The data, and source code, along with frequently updated dashboard monitoring the R-value will be made open and available for public use.
\end{enumerate}

\section{Methodology}

\subsection{Effective Reproduction Number}
The real time reproduction number is estimated using Bayesian approach, assuming the new number of daily cases satisfies the Poisson paradigm. The work\cite{kevin_sys} done by Kevin Systrom at the US state level is replicated here. Systrom used the modified version of a solution created by \cite{bettencourt2008real} to estimate a real time reproduction number.
As with changing conditions (behavior of people, government policies, etc), the value of $R_t$ changes. The effective reproduction number depends on yesterday’s (or previous) reproduction number and number of daily new cases. \cite{bettencourt2008real} use Bayes’ rule to update the real time value of $R_t$ from the number of daily new cases and prior value of reproduction number.

The new number of cases are seen everyday. This number of new cases says us something about the tranmissibility. Also, the $R_t$ value of today has relation with $R_{t-1}$ value of yesterday, and every previous value of $R_{t-m}$.

\cite{bettencourt2008real} use Bayes' rule to update the true value of $R_t$ based on the number of new cases daily.

Mathematically,

$$ P(R_t|k)=\frac{P(k|R_t)\cdot P(R_t)}{P(k)} $$

So, if we see $k$ new cases, the distribution of $R_t$ is equal to the likelihood of seeing $k$ new cases given $R_t$ times the prior beliefs of the value of $P(R_t)$ without the data divided by the probability of seeing this many cases in general.

Now, every day that passes, we use yesterday's prior $P(R_{t-1})$ to estimate today's prior $P(R_t)$. The distribution of $R_t$ is assumed to be a Gaussian centered around $R_{t-1}$, i.e. $P(R_t|R_{t-1})=\mathcal{N}(R_{t-1}, \sigma)$, where $\sigma$ is a hyperparameter.

\subsubsection*{Choosing a Likelihood Function $P\left(k|R_t\right)$}

A likelihood function function says how likely we are to see $k$ new cases, given a value of $R_t$.

We model the probability of seeing $k$ new cases according to Poisson distribution, with arrival rate $\lambda$ equal to number of new cases each day.

$$P(k|\lambda) = \frac{\lambda^k e^{-\lambda}}{k!}$$

\subsubsection*{Connecting $\lambda$ and $R_t$}

The connection between $R_t$ and $\lambda$ is given in the paper as:

$$ \lambda = k_{t-1}e^{\gamma(R_t-1)}$$

where $\gamma$ is the reciprocal of the serial interval. The serial interval is about 7 days according to CDC. As we know the number of new cases on the previous day, we can reformulate the likelihood function as a Poisson parameterized by fixing $k$ and varying $R_t$.

$$ \lambda = k_{t-1}e^{\gamma(R_t-1)}$$

$$P\left(k|R_t\right) = \frac{\lambda^k e^{-\lambda}}{k!}$$
\section{Limitations}
\label{sec:headings}

\subsection{Google mobility data}
There are blind spots in alternative data, in particular data coming from mobile phones. Lower smartphone penetration rate among older people, and rural population, may not give a complete picture of the mobility. Additionally, the Google mobility report states that their data comes only from android smartphone users who allowed the device to track their location. 

\subsection{Oxford Government Response tracker}
Oxford Coronavirus Government response tracker does not aim to measure the appropriateness or effectiveness of a country’s response. So a higher index should not be interpreted as the efficacy or effectiveness of the policy. 

\section{Background}
After several months of relatively low COVID cases in Nepal , COVID-19 cases began to rapidly spike in mid-April 2021 following a steep upwards trajectory (Figure 1).

\textcolor{black!65}{Figure 1: Cases, stringency index\footnote{“The Stringency Index is an aggregate score/composite measure made up of a particular combination of policy indicators/response metrics from the codebook and their values (for the Stringency Index these are C1-8 and H1). The OxCGRT aggregates these policy indicator values into a common “Stringency Index” that runs from 0 -100.”- www.bsg.ox.ac.uk}, and residential mobility\footnote{Stay at home requirements: 0 - no measures , 1 - recommend not leaving house , 2 - require not leaving house with exceptions for daily exercise, grocery shopping, and 'essential' trips , 3 - require not leaving house with minimal exceptions (eg allowed to leave once a week, or only one person can leave at a time, etc) , Blank - no data}}

\includegraphics[scale=0.45]{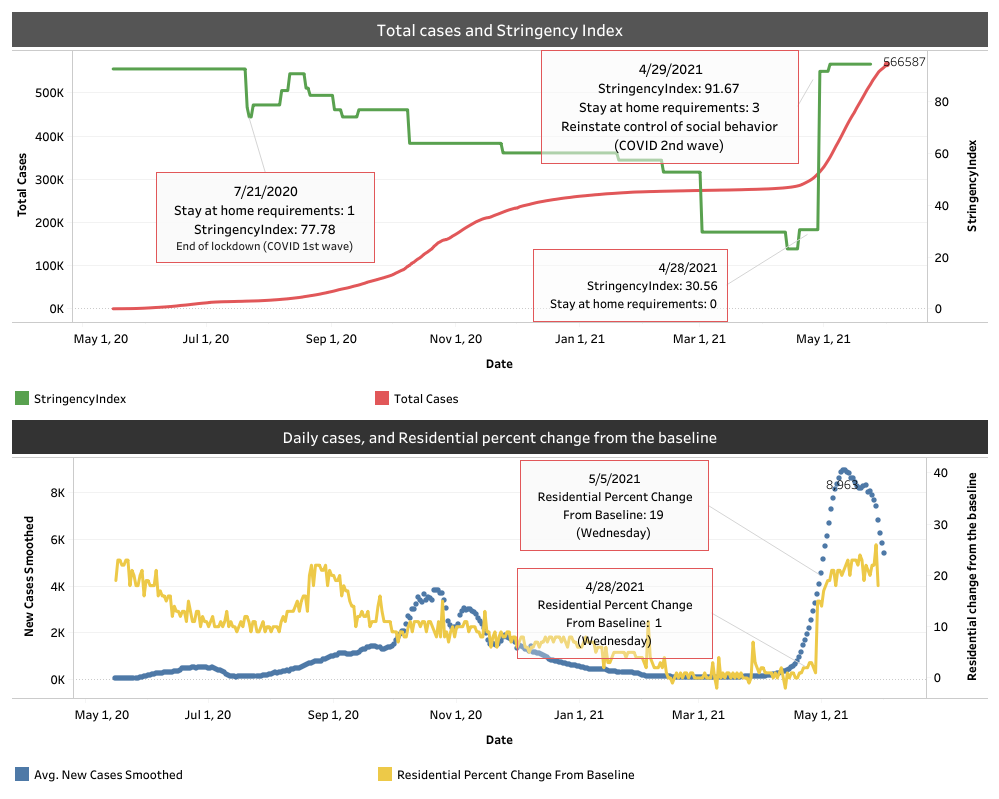}

\textcolor{black!65}{Source: (Our World in Data); (Google LLC, 2020); (University of Oxford - Blavatnik School of Government, 2020)}

\pagebreak

It was on 29th April 2021 after the cases started to surge, the transmission rate shooted upto 2.480 on 21st April - highest observed by the country till date (Figure 2). Nepal reinstated the control of social behavior, the stringency index which measures the severity of Government response increased from 30.56 to 91.67, as the Government increased the stay home requirements from 0 to 3 i.e from “no measures” to “require not leaving house with minimal exceptions”. People spent more time at home, and there was a spike in “residential percentage change from the baseline\footnote{All lines represent a 7-day moving average. Baseline values were established using a median of the corresponding day of the week from the period between January 3 and February 6, 2020.}” from 1\% on 28th April 2021 to 19\% on 5th May 2021.

\textcolor{black!65}{Figure 2: R-value and stringency index}

\includegraphics[scale=0.45]{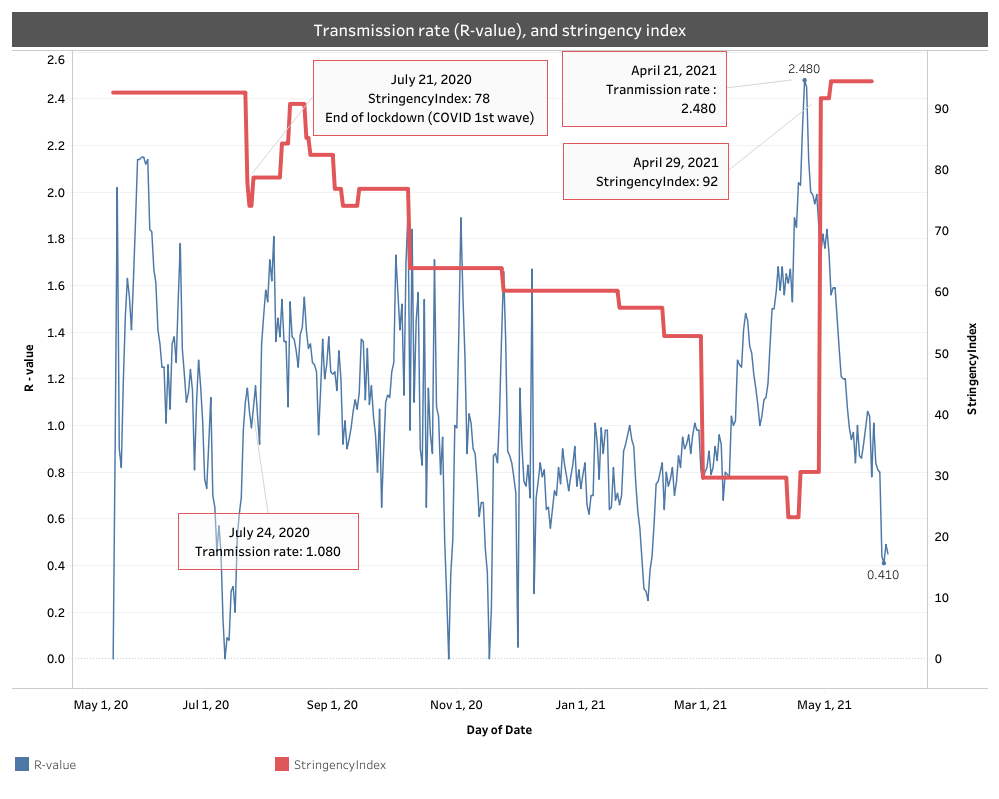}

\pagebreak

Grocery and pharmaceuticals - generally experiencing high mobility, also  recorded slump in the movement on 5th May 2021 and thereon (Figure 3). On Wednesday 28th  April  2021, the change in mobility in grocery and pharmaceuticals in Nepal was 81 percent, whereas on Wednesday 5th May 2021 it receded to -21 percent. This can also be interpreted as “people were 21 percent less likely to be in grocery and pharmaceuticals on Thursday 5th  May” than they were in the baseline (median of the corresponding day between 3rd January and 6th February). Similarly, the people were 81 percent more likely to be in grocery and pharmaceuticals on 28th April 2021\footnote{Because of the privacy, google did not release the absolute number.}. This was a result of restriction on movement as the Government reduced the opening hours of Grocery. The index of restriction on internal movement went up to 2. 

\textcolor{black!65}{Figure 3: Grocery and pharma percent change from baseline, and restriction on internal movement}

\includegraphics[scale=0.55]{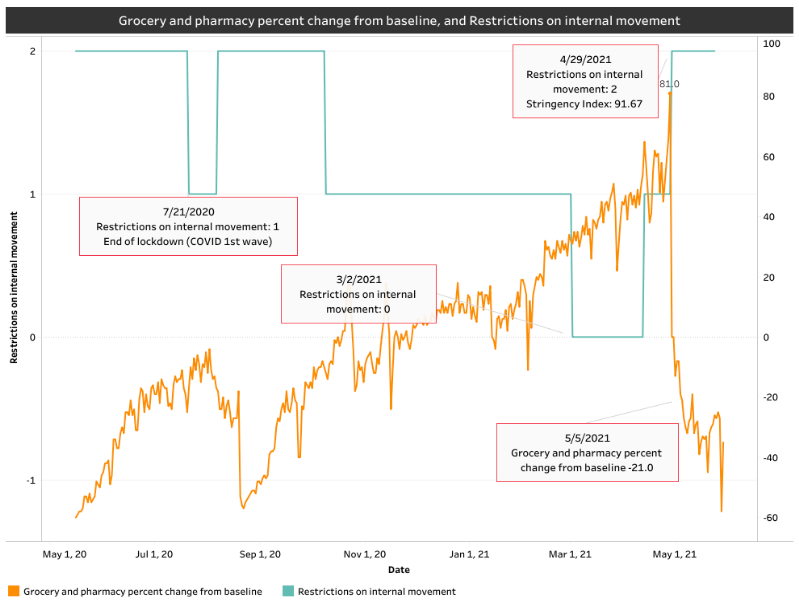}

\textcolor{black!65}{Source: (Google LLC, 2020)\cite{Google_mobility}; (University of Oxford - Blavatnik School of Government, 2020)}

\pagebreak

The movement in transit and stations changed from 56 percent in 28th April to -41 percent in 5th May from the baseline, approximately 43 percent points increase. The closure of public transport, and restrictions in internal movement imposed by the Government decreased the frequency of visits in the transit stations such as bus parks and airports. However, some movements were allowed with travel passes (Figure 4). 

\textcolor{black!65}{Figure 4: Transit and station}

\includegraphics[scale=0.55]{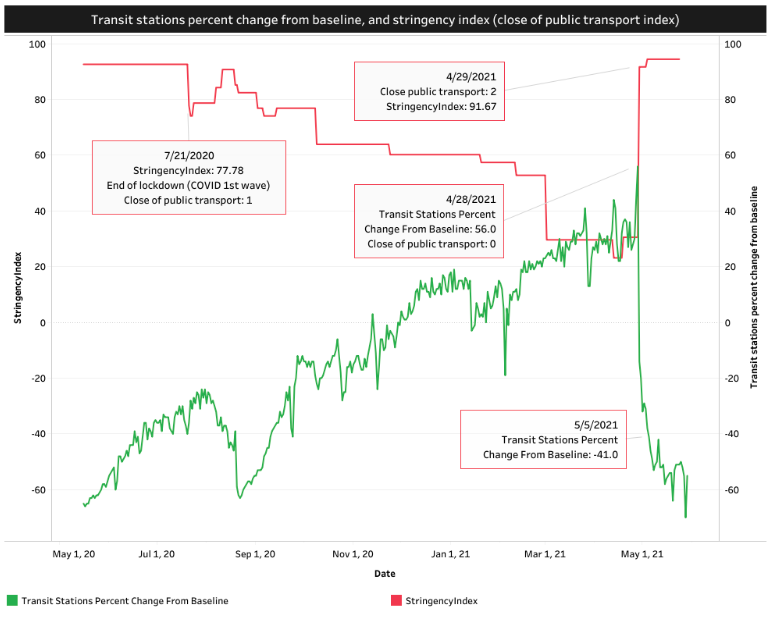}

\textcolor{black!65}{Source: (Google LLC, 2020); (University of Oxford - Blavatnik School of Government, 2020)}

\pagebreak

In the Google mobility report, the baseline for the weekend is the median of weekends falling between January 3 and February 6. As the weekend visits get closer to normal value the relative change becomes smaller, as such we see recurrent spikes in workplace mobility (Figure 5).  In Nepal, while most of the corporate offices, IT companies, NGOs and INGOs,  switched to work from home\footnote{https://kathmandupost.com/art-culture/2020/03/22/covid-19-and-the-shift-to-remote-working}, there were banks and financial institutions which continued to operate as per the regulatory instructions\footnote{https://thehimalayantimes.com/business/banks-to-remain-open-in-kathmandu-valley-during-prohibitory-period} and remained open even during the lockdown with limited staff and reduced hours of operations. 

\textcolor{black!65}{Figure 5: Workplace and mobility\footnote{\textbf{Close public transport} 0: No measures; 1: Recommended closing (or significantly reduce volume/route/ means of transport available)\\
\textbf{School closing} 0: No measures; 1: Recommended closing; 2: Required closing (only some levels or categories, eg just high school, or just public schools; 3: require closing all levels.}}

\includegraphics[scale=0.6]{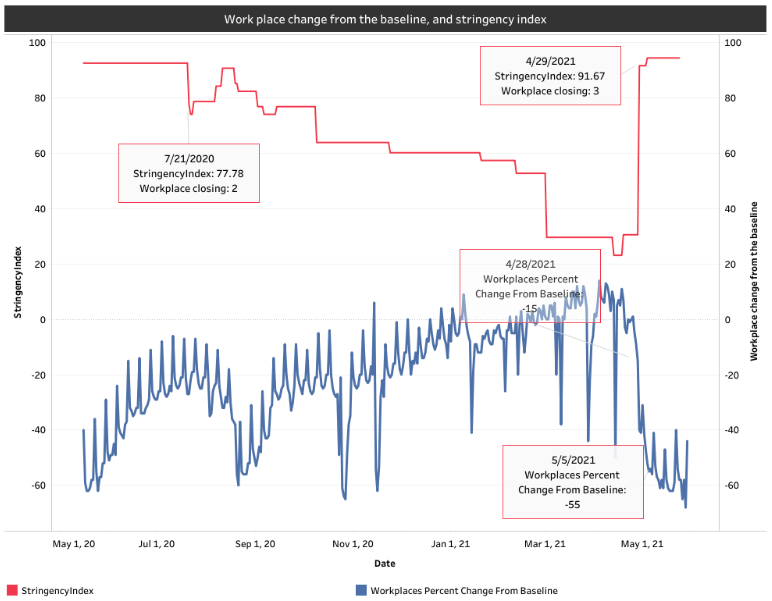}

\textcolor{black!65}{Source: (Google LLC, 2020); (University of Oxford - Blavatnik School of Government, 2020)}

\pagebreak

Furthermore, the schools closed and stopped in-person learning (Figure 5). Since most of the private school and students opted for distant learning\footnote{https://www.nepalitimes.com/banner/lockdown-gives-distance-learning-a-boost-in-nepal/}, the alternate data from Google trends reveal that the interest to download “zoom” and “google meet” - video communication software exceeded even the interest in “songs”, “music”, and “games”  during the lockdown period (Figure 6).

\textcolor{black!65}{Figure 6: Google trends – interest over time\footnote{Explore more here :\\ https://trends.google.com/trends/explore?geo=NP\&q=Zoom\%20download,Download\%20music,Download\%20songs,Download\%20games,\\Google\%20meets}}

\includegraphics[scale=0.6]{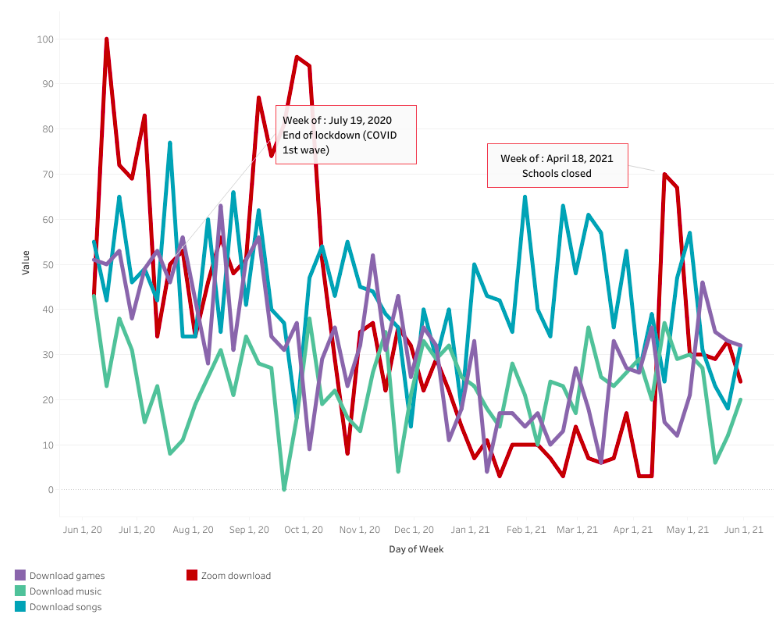}

\textcolor{black!65}{Source: Google trends}

\section{R-value and control measures}
The Government of Nepal has been quick to close the schools (Figure 7). The stay at home requirements, restrictions on gathering and cancellation of public events were made stringent only after the reproduction rate reached 2.48 on 21st April 2021. 

The scientific literature supports that the restriction on social behavior can work to break the chain of infection. It is accepted that more strict and timely restrictions have significant effects than slower, weaker ones. The radical control of social behavior has helped the Government of Nepal drop R down to about 0.450. But while it works on an average, there is no guarantee these measures will always work. Evidence from countries like Peru which suffered rising disease despite restrictive policies, reinforcing the fact that compliance and trust are also key to effectiveness\footnote{https://theconversation.com/what-we-learned-from-tracking-every-covid-policy-in-the-world-157721}. 

Moreover, without vaccinating the majority of the population the reproduction number could shoot-up if people start losing patience with restrictions, or the Government eases the measures (Figure 7). 
As such there is an unprecedented challenge in front of scientists and policy makers in Nepal and around the world, the criteria for policy adjustments are unknown . Is cancelling the public events doing the heavy lifting, or closing schools? How much economic support would help pacify the situation? How far should the vaccination coverage be stretched ? There is an ample scope for future research to answer this policy dilemma to keep the pandemic control but at an acceptable economic and social cost\footnote{ https://www.nytimes.com/2020/04/06/opinion/coronavirus-end-social-distancing.html}.

\textcolor{black!65}{Figure 7: R-value and Government Policies\footnote{\textbf{Vaccination policy}: 0 - No availability , 1 - Availability for ONE of following: key workers/ clinically vulnerable groups (non elderly) / elderly groups , 2 - Availability for TWO of following: key workers/ clinically vulnerable groups (non elderly) / elderly groups , 3 - Availability for ALL of following: key workers/ clinically vulnerable groups (non elderly) / elderly groups , 4 - Availability for all three plus partial additional availability (select broad groups/ages) , 5 - Universal availability
}}

\includegraphics[scale=0.6]{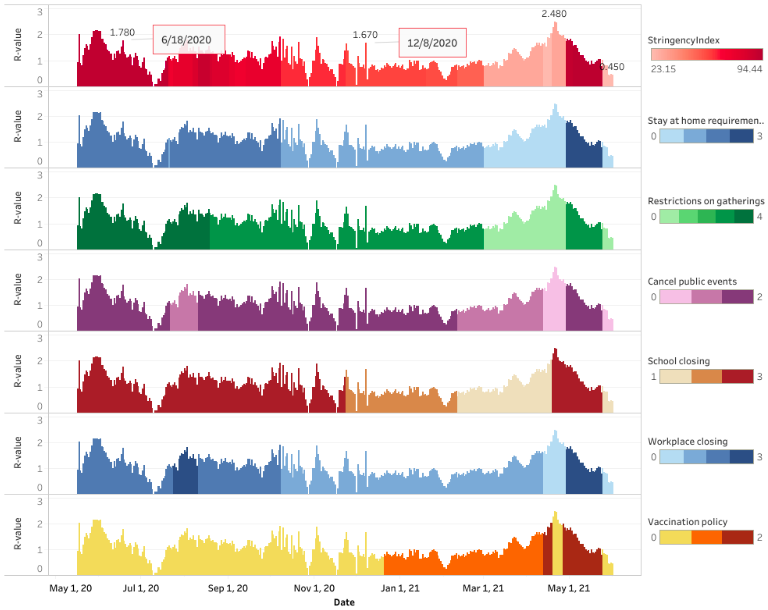}

\section{Conclusion}

As the country is experiencing the data challenges during the COVID crisis, there is an urgent need to gear up efforts to use alternative data. The high frequency readout of the situation from alternative data can be useful for the Government to get the real time assessment of the R-value, and understand social behavior. 

The Government of Nepal has been responding with the radical control of social behaviors and it has resulted in decreased mobility.  Nepal’s COVID reproduction rate (R) is now down to 0.41, but if people start losing patience with restrictions or if the Government relaxes the control, R could quickly rise again. As such the vaccinating majority of the population becomes imperative.

As the government tentatively eases lockdown restrictions around the world, it will be monitoring R very cautiously. The future research could be on finding the optimum policy response for the Government to tune their interventions quickly enough to stay ahead of the outbreak trajectory. 

\section*{Acknowledgments}
Authors would like to acknowledge the  guidance received from Hiroki Uematsu, Senior Economist at the World Bank.

\bibliographystyle{unsrt}  
\bibliography{references.bib}

\end{document}